\documentclass[aps,pra,reprint,superscriptaddress,nofootinbib,longbibliography]{revtex4-2}

\usepackage[T1]{fontenc}
\usepackage[utf8]{inputenc}
\usepackage{amsmath,amssymb,mathtools}
\usepackage{booktabs}
\usepackage{siunitx}
\usepackage{hyperref}
\usepackage{microtype}
\usepackage{enumitem}
\usepackage{array}

\hypersetup{
  colorlinks=true,
  linkcolor=blue,
  citecolor=blue,
  urlcolor=blue
}

\newcommand{\MI}{I}
\newcommand{\TV}{\mathrm{TV}}
\newcommand{\bits}{\{0,1\}}
\newcommand{\ftrip}{f(123)}
\newcommand{\gtrip}{g(123)}

\begin{document}

\title{Empirical Falsification of Pairwise-Only Explanations \\ for an Engineered Parity Benchmark on a 133-Qubit Superconducting Processor}

\author{Petr Sramek}
\email{p.sramek@whytics.com}
\affiliation{Whytics, Cambridge, MA, USA}

\date{March 2026}

\begin{abstract}
Scalable quantum characterization and error-mitigation workflows often rely on the assumption that relevant device noise and readout contamination can be adequately captured by low-weight, predominantly pairwise interactions. We report a compact hardware experiment designed to operationally distinguish pairwise-only explanations from irreducible triplet-order predictive structure. The A1/A1b protocol implements a parity-structured binary label on a 133-qubit IBM superconducting processor (\texttt{ibm\_torino}) and analyzes the resulting data through a classical M\"obius decomposition of subset mutual informations. In the A1 baseline, we observe a macroscopic triplet correlation of $\ftrip=\num{0.72609}$ bits ($p\le 1.0\times10^{-4}$, permutation floor). In the strict A1b loophole-reduction follow-up, role-symmetry averaging sharply suppresses singleton leakage, modestly reduces pairwise mismatch, and preserves a large irreducible triplet term of $\ftrip=\num{0.56521}$ bits. Crucially, a principled pairwise maximum-entropy baseline consistent with the empirical 1- and 2-body marginals implies only $\ftrip\approx\num{6.6e-06}$ bits, in strong contradiction with the observed hardware data. On A1b, a classifier built exclusively from pairwise features reaches only \num{0.617} held-out accuracy (chance \num{0.5}), whereas a triplet-inclusive model reaches \num{0.910}. These results provide a concise, open-data demonstration that pairwise benchmarking proxies can be fundamentally blind to higher-order contextual structure in present-day superconducting experiments.
\end{abstract}

\maketitle

\section{Introduction}

Modern Quantum Characterization, Verification, and Validation (QCVV) pipelines and noise-tailoring protocols often assume that the practically relevant structure of a multiqubit device can be well approximated by single-variable errors and localized pairwise interactions \cite{Harper2020, Proctor2022}. On superconducting architectures, this often appears as the assumption that residual $ZZ$-crosstalk, spectator dephasing, and readout assignment errors decay rapidly with topological distance and can be effectively modeled by two-body Pauli-Lindblad generators \cite{Berg2022}. 

However, that assumption fails dramatically for macroscopic parity-like structure, where all one- and two-body marginals can be completely uninformative (maximally mixed) while the full joint state remains strongly correlated with the distinguishing context \cite{SramekObservable2026}. The present paper turns that observation into a deliberately minimal hardware falsification experiment.

Instead of deploying full multitime process-tensor tomography or related non-Markovian reconstructions \cite{White2022, Milz2021}, which scale exponentially, we isolate this effect using classical subset-lattice metrics evaluated directly on measurement readouts. One defines a cumulative information observable on readout subsets and then applies M\"obius inversion to isolate irreducible contributions by interaction order \cite{Shannon1948, Rota1964}. A1/A1b is the minimal order-3 synthetic benchmarking harness: rather than inferring high-order structure from an unprogrammed many-body bath, it engineers a measurement distribution in which the triplet channel is the intended carrier of context information, and then asks whether that structure survives real hardware noise without collapsing into standard pairwise proxies.

The core falsifiable question is narrow and operational: \emph{Can the context-predictive structure in the measured outcomes be explained by single-bit and pairwise statistics alone?} If yes, then the observed triplet signal should collapse under a principled pairwise maximum-entropy model. If no, a large positive triplet M\"obius term should remain, and pairwise-only surrogates should operationally fail.

\section{Theory: Cumulative Information and Irreducible Triplet Structure}

Let $Y\in\bits$ denote the binary macroscopic context label and let $X_1,X_2,X_3\in\bits$ denote the observed output bits of the probe register. For every nonempty subset $S\subseteq\{1,2,3\}$, define the cumulative observable
\begin{equation}
  g(S) \equiv \MI(Y;X_S),
\end{equation}
where $X_S$ is the tuple formed by the variables indexed by $S$. This is a standard Shannon-entropic choice for discrete multiscale validation, bounding multi-body correlations from classical measurement records \cite{McGill1954}.

On the Boolean subset lattice, M\"obius inversion gives intrinsic contributions $f(S)$ satisfying
\begin{equation}
  g(S)=\sum_{T\subseteq S} f(T), \qquad f(S)=\sum_{T\subseteq S} \mu(T,S)g(T),
\end{equation}
with $\mu(T,S)=(-1)^{|S|-|T|}$ \cite{Rota1964}. For three variables, the irreducible triplet term is
\begin{align}
  f(123) =&~ g(123)-g(12)-g(13)-g(23) \nonumber \\
  &+g(1)+g(2)+g(3),
\end{align}
using the convention $g(\varnothing)=0$. Operationally, $f(123)$ measures the information about $Y$ that is available \emph{only} from the joint observation of all three variables and is strictly not attributable to any singleton or pair subset. This is the order-3 blade edge of the experiment.

\subsection*{Ideal Parity Construction}
Consider the idealized label-conditional distributions
\begin{align}
  Y=0 &: X_1\oplus X_2\oplus X_3 = 0,\\
  Y=1 &: X_1\oplus X_2\oplus X_3 = 1,
\end{align}
with the valid strings for each label taken uniformly. Then each single bit is unbiased, each pair is uniform over its four outcomes, and therefore
\begin{equation}
  \MI(Y;X_i)=0,\qquad \MI(Y;X_iX_j)=0.
\end{equation}
However, the parity of the full triple determines the label exactly, yielding
\begin{equation}
  \MI(Y;X_1X_2X_3)=1~\text{bit}, \qquad f(123)=1~\text{bit}
\end{equation}
for the noiseless model. This is the sharp pairwise-vs-triplet separation that the A1/A1b harness attempts to realize on hardware.

\section{Experimental Design}

\subsection{Platform and Execution Model}
Both experiments were executed on IBM Quantum \texttt{ibm\_torino} (a 133-qubit processor with tunable couplers). We utilize deliberately shallow circuit families so that the intended structure is easy to interpret and standard depth-related coherent errors remain bounded. 

The basic A1 configuration uses 16 circuits total, corresponding to two labels and eight measurement-twirl masks, with 512 shots per circuit for 8192 total shots. The A1b follow-up expands this to 48 circuits by adding three role-symmetry variants, again at 512 shots per circuit for 24,576 total shots. In both cases, transpiled depth remained approximately 9, and the two-qubit gate budget remained minimal.

\subsection{A1 Baseline and A1b Loophole Reduction}
A1 is the minimal hardware existence proof. Its purpose is to test whether a parity-structured signal can produce a large irreducible triplet term despite real-device leakage in low-order marginals. Measurement twirling is used to reduce systematic readout asymmetries (hardware job: \texttt{d55enm1smlfc739gmhb0}).

A1b addresses the strongest hardware criticism of A1: a single qubit might play a privileged physical role and leak label information through static hardware asymmetries (e.g., specific readout resonator imbalances). To suppress this loophole, A1b rotates the parity-accumulator role among physical qubits $\{0,1,2\}$ while retaining the same parity target and measurement twirling (hardware job: \texttt{d55f9rvp3tbc73aoa860}). A1b therefore cleanly tests whether the discriminative structure is genuinely triplet-order rather than a byproduct of single-site bias.

\section{Data Analysis and Statistical Tests}

The measured data are represented as empirical distributions $\hat p(x_1,x_2,x_3\mid Y=y)$ aggregated across the relevant circuits. All mutual informations are computed via the discrete plug-in estimator, and the triplet term is obtained by direct inclusion--exclusion.

Uncertainty is quantified with bootstrap resampling over shots within each label (5,000 replicates) to form 95\% confidence intervals for $f(123)$. Statistical significance is quantified by a label-shuffle permutation test (10,000 shuffles). Because this test is Monte Carlo limited, floor-level significances are reported as $p\le 1.0\times 10^{-4}$ when no shuffled replicate reaches the observed statistic.

To connect this structural statistic to QCVV operational predictability, we compare two classical decoders on a stratified 50/50 train/test split: a pairwise-features model and a triplet-inclusive model. Because the label is balanced by construction, chance accuracy is \num{0.5}. Finally, A1b evaluates a principled pairwise-only maximum-entropy (Ising) model constrained to match the empirical 1- and 2-body marginals. If the observed triplet effect on the hardware were merely a dressed-up pairwise artifact, this surrogate would naturally reproduce it.

\section{Results}

\subsection{A1 Baseline Results}
A1 yields a massive and highly significant triplet signal. The maximum label-conditioned marginal mismatch is \num{0.01367}, and the maximum pairwise total variation (TV) distance is \num{0.03589}. Against that background, the cumulative triplet mutual information is $\gtrip=\num{0.72748}$ bits and the irreducible triplet term is $\ftrip=\num{0.72609}$ bits (95\% CI $[\num{0.70581},\num{0.74619}]$, permutation floor $p\le 1.0\times10^{-4}$). On the test split, the pairwise-features model reaches \num{0.622} accuracy, whereas the triplet-inclusive model reaches \num{0.953}. Thus, even with measurable hardware mismatch, the signal remains overwhelmingly parity-like.

\subsection{A1b Loophole-Reduction Results}
A1b sharply suppresses singleton leakage while only modestly reducing pairwise mismatch. The maximum absolute marginal difference drops from \num{0.01367} (A1) to \num{0.00326}, while the maximum pairwise TV distance changes only slightly from \num{0.03589} to \num{0.03410}. Despite this strict symmetry averaging, the triplet term remains robust: $\ftrip=\num{0.56521}$ bits (95\% CI $[\num{0.55352},\num{0.57738}]$, permutation floor $p\le 1.0\times10^{-4}$).

Low-order information leakage in A1b is extremely small in information-theoretic terms: the maximum singleton mutual information is $\num{7.65e-06}$ bits and the maximum pair mutual information is $\num{8.42e-04}$ bits. Operationally, the pairwise-features classifier reaches only \num{0.617} accuracy, whereas the triplet-inclusive model reaches \num{0.910}. 

\subsection{Pairwise-Only World Refutation}
The strongest A1b result is the pairwise maximum-entropy baseline. Under the fitted pairwise-only surrogate, the Bayes accuracy is \num{0.612} and the model-implied triplet term is approximately $\num{6.6e-06}$ bits, effectively zero. By contrast, the observed hardware data yield $\ftrip=\num{0.56521}$ bits. Thus the structural observable itself---not merely a particular classifier choice---rules out the claim that the measured outcomes are explainable by single-bit and pairwise statistics alone. Matched held-out classifier accuracies are reported in Table \ref{tab:headline}; Table \ref{tab:surrogate} focuses on the structural comparison.

\begin{table}[t]
  \centering
  \caption{A1 and A1b headline results on \texttt{ibm\_torino}. Pair/Trip accuracies are matched held-out test accuracies; chance level is \num{0.5}.}
  {\setlength{\tabcolsep}{3.5pt}
  \scriptsize
  \begin{tabular}{@{}lccccc@{}}
    \toprule
    Variant & \shortstack{Circuits\\/ Shots} & \shortstack{Max\\$|\Delta|$ marg.} & \shortstack{Max\\$\TV$ pair} & \shortstack{$f(123)$\\(bits)} & \shortstack{Acc.\\(Pair/Trip)} \\
    \midrule
    A1  & 16 / 8192  & 0.01367 & 0.03589 & 0.72609 & 0.622 / 0.953 \\
    A1b & 48 / 24576 & 0.00326 & 0.03410 & 0.56521 & 0.617 / 0.910 \\
    \bottomrule
  \end{tabular}
  }
  \label{tab:headline}
\end{table}

\begin{table}[t]
  \centering
  \caption{A1b pairwise-only surrogate versus observed hardware. Because the surrogate matches the empirical 1- and 2-body marginals by construction, the table emphasizes the structural quantity $f(123)$; matched held-out classifier accuracies appear in Table \ref{tab:headline}.}
  {\setlength{\tabcolsep}{3pt}
  \scriptsize
  \begin{tabular}{@{}lcc@{}}
    \toprule
    Quantity & Pairwise Surrogate & Observed A1b \\
    \midrule
    $f(123)$ magnitude & $\approx \num{6.6e-06}$ bits & \num{0.56521} bits \\
    Max 1-body MI & matched & $\num{7.65e-06}$ bits \\
    Max 2-body MI & matched & $\num{8.42e-04}$ bits \\
    \bottomrule
  \end{tabular}
  }
  \label{tab:surrogate}
\end{table}

\section{Discussion}

The experiment falsifies, within the measured statistical uncertainty, the statement that \emph{all predictive structure present in these engineered measurement outcomes is explainable by single-bit and pairwise statistics.} The M\"obius analysis isolates a large, highly significant irreducible triplet term, and A1b shows that the best pairwise-only world consistent with the empirical low-order statistics predicts essentially no triplet channel at all.

\subsection{Implications for Quantum Error Mitigation}
Standard error mitigation techniques, such as Probabilistic Error Cancellation (PEC) and pairwise Randomized Benchmarking (RB), rely on learning sparse noise profiles from low-weight marginals \cite{Berg2022}. The A1/A1b falsification demonstrates that an engineered context can remain hidden within the blind spots of these techniques. If similar higher-order correlated structure arises naturally during unprogrammed multi-qubit execution (e.g., through parasitic measurement crosstalk), pairwise-only mitigation strategies could miss or misattribute it. A1/A1b therefore establishes a controlled ground-truth harness for testing whether advanced QCVV tools can detect irreducible order-3 structure.

\subsection{Limitations and Scope}
Several limitations are important. First, all mutual informations are plug-in estimates from finite data, so very small low-order quantities can retain mild finite-sample bias even though the headline triplet effect is far larger. Second, A1b sharply suppresses singleton leakage but only modestly reduces pairwise mismatch; it is a loophole-reduction control, not a full pairwise-calibration procedure. Third, the present study probes an engineered order-3 benchmark on one device configuration and calibration day, so it should be read as an operational falsifier for pairwise-only explanations in this setting rather than a universal statement about ambient hardware noise. Finally, the present manuscript validates order-3 structure only; extending the same logic to higher-order structure will require larger benchmark families.

\subsection{What is Not Claimed}
This paper does \emph{not} claim direct identification of a native three-body Hamiltonian, nor a device-wide bound on irreducible tripartite error rates, nor a universal statement about all IBM hardware noise models. The claim is narrower and operationally relevant to QCVV: in this engineered parity benchmark, the measured contextual structure survives readout and crosstalk noise so robustly that its informational signature remains irreducibly triplet-order. Consequently, benchmarking suites that rely exclusively on pairwise total-variation checks or pairwise-feature decoders are fundamentally blind to this class of contextual correlation.

\section{Reproducibility and Data Availability}
The supplementary Zenodo data bundle accompanying this manuscript \cite{SramekA1_2025} includes the machine-readable evaluation files, pre-transpile circuit descriptions, transpiled QASM, qubit mappings, and calibration snapshots. The hardware execution identifiers on IBM Quantum are \texttt{d55enm1smlfc739gmhb0} (A1) and \texttt{d55f9rvp3tbc73aoa860} (A1b).

\end{document}